\documentclass[a4paper,oneside,reqno,12pt]{amsart} 
\usepackage{amsxtra}  

\def\Lcal{\mathcal{L}}

\def\Pcal{\mathcal{P}}

\def\Ucal{\mathcal{U}}

\newcommand{\re}{\mathrm{e}}            
\newcommand{\ri}{\mathrm{i}}            

\def\d{\delta}
\def\eps{\varepsilon}
\def\la{\lambda}

\def\phi{\varphi}

\def\La{\Lambda}
\def\tr{\mathop{\hbox{\rm tr}}\nolimits}

\def\dd{\partial}
\def\Ref#1{(\ref{#1})}
\def\tilde{\widetilde}
\def\endproof{\hfill\rule{2mm}{2mm}}

\def\bra#1{\bigl<#1\bigr|}
\def\ket#1{\bigl|#1\bigr>}

\def\ft{\tilde{f}}
\def\gt{\tilde{g}}

\def\ft{\tilde{f}}
\def\gt{\tilde{g}}

\def\Phit{\tilde{\Phi}}
\def\phit{\tilde{\varphi}}
\def\endproof{\hfill\rule{2mm}{2mm}}
\def\?{(?)\marginpar[\hfill?|]{|?}}

\def\beq{\begin{equation}}
\def\eeq{\end{equation}}
\def\be{\begin{equation*}}
\def\ee{\end{equation*}}
\def\bmat{\left(\begin{array}}
\def\emat{\end{array}\right)}
\newcounter{problem}
{\addtocounter{problem}{1}{\bf Çàäà÷à \theproblem.\marginpar[{\hfill\maltese}]{{\maltese}}}\em }{}

 \makeatletter
 \@addtoreset{equation}{section}
 \makeatother
 
\begin{document}

\begin{center}
\Large\bf
Bispectrality for the quantum open Toda chain
\end{center}
\vskip1cm

\begin{center}
\large\bf
E.\ K.\ Sklyanin
\end{center}
\vskip1cm
\textbf{Abstract.}
An alternative to Babelon's (2003) construction of dual variables for the quantum open Toda chain
is proposed that is based on the $2\times2$ Lax matrix and the corresponding quadratic
$R$-matrix algebra. 

\section{Introduction}

The term {\em bispectrality} \cite{DuiGr1986} refers to a situation when a function
$f(x,\la)$ depending on two variables $x$ and $\la$ simultaneously satisfies two 
spectral problems: one being a differential or finite-difference equation $H(x,\dd_x) f=\la f$ in $x$, with $\la$ playing the role of the spectral parameter, and the dual one, $\tilde{H}(\la,\dd_\la) f= x f$
in $\la$, with $x$ being the spectral parameter. The function $f(x,\la)$ can thus be considered as a kernel
of an integral operator providing expansion in the eigenfunctions of the operator $H$, or of $\tilde{H}$,
depending on the point of view.

In the simplest examples, the bispectrality
is a manifestation of the contiguity relations for the hypergeometric function.

In the context of multivariate quantum integrable systems, $x$ and $\la$ become 
sets of $N$ variables, and the two spectral problems become spectral problems for commuting
quantum Hamiltonians. 
The bispectral formulation has been found for the quantum open Toda chain \cite{Bab2003,Bab2004},
the quantum Calogero-Moser system and its generalisations \cite{Chal2000},
the Knizhnik-Zamolodchikov equation \cite{TarVar2002}, Gaudin models \cite{MTV2008}, 
to name a few examples.

The bispectrality seems to be a common property enjoyed by multidimensional hypergeometric equations,
among their other characteristic properties, like existence of integral representations for the eigenfunctions
with the kernel expressed in terms of elementary functions, raising/lowering operators etc.


The purpose of this letter is to revise the case of the quantum open Toda chain
and to suggest a few improvements to Babelon's \cite{Bab2003,Bab2004} original construction of the dual variables for that model. Like Babelon, we start with the construction for the classical (non-quantum) case.
However, unlike \cite{Bab2003}, we do not consider the periodic case that leads to difficult algebraic-geometric
constructions for hyperelliptic curves. Instead, we start directly with the open Toda chain and use the observation of \cite{KV2000} that the corresponding spectral curve is a rational algebraic curve.
The resulting formulae are the same as in \cite{Bab2003} but the derivation is simplified drastically.

In \cite{Bab2003,Bab2004} the formulae for the quantum case were conjectured based on the results for the classical case. However, the proof of the conjectured formulae given in \cite{Bab2003,Bab2004} does not follow the classical construction and is based instead on the integral representations for the eigenfunctions found in
\cite{KharLeb2001}. Our derivation has an advantage that it is completely parallel to the classical case
and is considerably simpler that in the pioneering papers \cite{Bab2003,Bab2004}.
Besides, being based on the $R$-matrix algebra for the monodromy matrix, it allows, in principle, generalisations to other integrable models.

\section{Description of the model}.

We start with a discussion of the classical case. In what follows we use the notation
of \cite{EKS16,EKS53}.

The open $n$-particle Toda chain
is described in terms of the canonical variables
\beq
  \{X_j,X_k\}=\{x_j,x_k\}=0, \quad \{X_j,x_k\}=\d_{jk},\quad
  j,k=1,\ldots,n.
\label{eq:Pb}
\eeq
and is characterised by the Hamiltonian
\beq
 H=\sum_{j=1}^n \frac12 X_j^2+\sum_{j=1}^{n-1} \re^{x_{j+1}-x_j}.
\label{eq:def-H}
\eeq

Defining local Lax matrices as
\beq
 \ell_j(u)=
     \begin{pmatrix}
     u+X_j & -\re^{x_j} \\
     \re^{-x_j} & 0 \end{pmatrix}
\label{eq:def-ell}
\eeq
we introduce the partial monodromy matrices
\begin{subequations}\label{defLjk}
\begin{align}
    L_{jk}(u)&\equiv \ell_j(u)\ell_{j-1}(u)\ldots\ell_{k+1}(u)\ell_k(u)
    =\begin{pmatrix}
     A_{jk}(u) & B_{jk}(u) \\
     C_{jk}(u) & D_{jk}(u)
   \end{pmatrix}, \qquad j\geq k,\\
   L_{jk}&\equiv \begin{pmatrix}1&0\\0&1\end{pmatrix},\qquad
   j<k,
\end{align}
\end{subequations}
and the complete monodromy matrix
\beq
 L(u)\equiv L_{n1}(u)=\ell_n(u)\ldots\ell_2(u)\ell_1(u)
 =\begin{pmatrix}
     A(u) & B(u) \\
     C(u) & D(u)
   \end{pmatrix}.
\label{eq:toda-Lll}
\eeq

To study the open Toda chain it is sufficient to work with the single column-vector $AC$:
\beq
    \begin{pmatrix}A(u) \\ C(u)\end{pmatrix}
    \equiv \begin{pmatrix}A_{n1}(u) \\ C_{n1}(u)\end{pmatrix}
    =L(u) \begin{pmatrix}1\\0\end{pmatrix},
\eeq
defined recursively:
\beq
    \begin{pmatrix}A_{j1}(u) \\ C_{j1}(u)\end{pmatrix}
    =     \begin{pmatrix}
     u+X_j & -\re^{x_j} \\
     \re^{-x_j} & 0 \end{pmatrix}
    \begin{pmatrix}A_{j-1,1}(u) \\ C_{j-1,1}(u)\end{pmatrix}, \quad
    \begin{pmatrix}A_{01}(u) \\ C_{01}(u)\end{pmatrix}
    = \begin{pmatrix}1\\0\end{pmatrix},
\eeq
or, componentwise
\begin{subequations}
\begin{align}
   A_{j1}(u)&=(u+X_j)A_{j-1,1}(u)-\re^{x_j}C_{j-1,1},\\
   C_{j1}(u)&=\re^{-x_j}A_{j-1,1}(u),
\end{align}
\end{subequations}
\beq
    A_{j1}(u)=(u+X_j)A_{j-1,1}(u)-\re^{x_j-x_{j-1}}A_{j-2,1}(u).
\eeq

The generating function of the integrals of motion of the open chain
\beq
 A(u)=u^n+H_1u^{n-1}+\ldots+H_n.
\label{eq:def-Hj}
\eeq
\beq
    H_1=X_1+\ldots+X_n, \qquad H_2=\frac12 H_1^2-H.
\eeq
can be considered as the limit as $\eps\rightarrow0$
of the quasiperiodic chain 
\beq
    t_\eps(u)=\tr LK_\eps=A(u)+\eps D(u), \qquad
    K_\eps=\begin{pmatrix}1&0\\0&\eps\end{pmatrix},
\eeq
\beq
    A(u)=\tr LK_0, \qquad
    K_0=\begin{pmatrix}1&0\\0&0\end{pmatrix},
\eeq
with the boundary condition (twist) determined by the matrix $K_\eps$,
when the hyperelliptic spectral curve
\beq
    \det\bigl(v-L(u)\bigr)=v^2-t_\eps(u)v+\eps=0
\eeq
degenerates into a rational one
\beq
    v=A(u).
\eeq

The observation that the spectral curve for the open Toda chain
is rational is due to \cite{KV2000} and plays the key role in our construction.
Working directly with the rational curve rather with the functions on the hyperelliptic curves
and then taking the limit $\eps\rightarrow0$ in the end, allows to simplify the derivations
considerably.

The local Lax matrices \Ref{eq:def-ell} satisfy 
the $r$-matrix Poisson brackets relations
\beq
    \{\overset{1}{\ell}(u_1),\overset{2}{\ell}(u_2)\}
    =[r(u_{12}),\overset{1}{\ell}(u_1)\overset{2}{\ell}(u_2)],\quad
    r(u)=\frac{\Pcal_{12}}{u},
\label{eq:Pb-ellell}
\eeq
where $u_{12}\equiv u_1-u_2$ and $\Pcal_{12}$ is the permutation operator
(see \cite{EKS16,EKS53} for the explanation of the notation).

The relations \Ref{eq:Pb-ellell} imply immediately the same relations for the monodromy
matrices $L_{jk}(u)$
\beq
    \{\overset{1}{L}_{jk}(u_1),\overset{2}{L}_{jk}(u_2)\}
    =[r(u_{12}),\overset{1}{L}_{jk}(u_1)\overset{2}{L}_{jk}(u_2)],
\label{eq:Pb-LL}
\eeq
including the case $L(u)\equiv L_{n1}(u)$. 
As a consequence,
\beq
    \{A(u_1),A(u_2)\}=\{C(u_1),C(u_2)\}=0,
\eeq
\beq
    \{A(u_1),C(u_2)\}=
    \frac{-A(u_1)C(u_2)+C(u_1)A(u_2)}{u_1-u_2}.
\eeq

The dual variables $(\la_j,\La_j)$, $j=1,\ldots,n$
are introduced by the equations
\begin{subequations}
\begin{alignat}{2}
    \la_j:&\qquad& A(\la_j)&=0,\\
    \La_j:&\qquad& \La_j&=C(\la_j),
\end{alignat}
\end{subequations}
and one can verify that
\beq\label{pb-lala}
   \{\La_j,\La_k\}=\{\la_j,\la_k\}=0,\qquad
   \{\La_j,\la_k\}=-\La_j\d_{jk}.
\eeq

The $AC$ vector is reconstructed in terms of the dual variables
through the interpolation formulae:
\begin{subequations}\label{AC-interpol}
\begin{align}
    A(u)&=\prod_{j=1}^n (u-\la_j),  \label{A-interpol} \\
    C(u)&=\sum_{j=1}^n \La_j \left( \prod_{k\neq j} \frac{u-\la_k}{\la_j-\la_k}\right). \label{C-interpol}
\end{align}
\end{subequations}

An alternative approach is based on the `large' $n\times n$ Lax matrix instead of the
$2\times2$ as above \cite{Fl1974,Man1974}.

The open chain is served by the Lax matrix without the spectral parameter
($x_{jk}\equiv x_j-x_k$):
\beq
\Lcal=\begin{pmatrix}
  -X_1 & 1 & \ldots & 0 & 0 \\
  \re^{x_{21}} & -X_2 & \ldots & 0 & 0 \\
  \ldots & \ldots & \ldots & \ldots & \ldots \\
  0 & 0 & \ldots & -X_{n-1} & 1 \\
  0 & 0 & \ldots & \re^{x_{n,n-1}} & -X_n
     \end{pmatrix},
\label{eq:def-LL}
\eeq

The spectral parameter for the quasiperiodic chain
is introoduced by adding two one-dimensional projectors
(rank 2 perturbation, \emph{ergo} hyperelliptic curve):
\begin{align}
\Lcal(v)&=\begin{pmatrix}
  -X_1 & 1 & \ldots & 0 & \eps v^{-1}\re^{x_{1n}} \\
  \re^{x_{21}} & -X_2 & \ldots & 0 & 0 \\
  \ldots & \ldots & \ldots & \ldots & \ldots \\
  0 & 0 & \ldots & -X_{n-1} & 1 \\
  v & 0 & \ldots & \re^{x_{n,n-1}} & -X_n
     \end{pmatrix}\notag\\
  &=\Lcal+v\ket{g}\bra{f}+\eps v^{-1}\ket{\gt}\bra{\ft},
\label{eq:def-LLv}
\end{align}
where
\beq
    \bra{f}\equiv(1,0,\ldots,0),\quad
    \bra{\ft}\equiv(0,\ldots,0,\re^{-x_n}),
\eeq

\beq
    \ket{g}=\begin{pmatrix}0\\\vdots\\0\\1\end{pmatrix},\qquad
    \ket{\gt}=\begin{pmatrix}\re^{x_1}\\0\\\vdots\\0\end{pmatrix}.
\eeq

The commuting Hamiltonians for the open chain are obtained as coefficients of the characteristic polynomial
\beq
     \det(u-\Lcal)=A(u),
\eeq
respectively, for the quasiperiodic chain,
\beq
     \det\bigl(u-\Lcal(v)\bigr) = -v^{-1}\det\bigl(v-L(u)\bigr).
\eeq

Introducing the adjunct matrix $\Ucal\equiv(u-\Lcal)\sphat=\det(u-\Lcal)(u-\Lcal)^{-1}$, 
we get four identities:
\begin{subequations}\label{4idents}
\begin{align}
    \bra{f}\Ucal\ket{g}&=\Ucal_{1n}=1,\\
    \bra{\ft}\Ucal\ket{g}&=\re^{-x_n}\Ucal_{nn}=C(u),\\
    \bra{f}\Ucal\ket{\gt}&=\re^{x_1}\Ucal_{11}=-B(u),\\
    \bra{\ft}\Ucal\ket{\gt}&=\re^{x_{1n}}\Ucal_{n1}=1.
\end{align}
\end{subequations}

\section{Solving  inverse problem}

{\bf Problem.} {\it Express the original variables $\re^{\pm x_j}$, $X_j$ 
in terms of the dual ones $\La_j$, $\la_j$.}

The problem was stated and the solution for $n=3$ was given by Kuznetsov \cite{Kuz2002} 
who made an important observation that 
$\re^{\pm x_j}$, $X_j$ are expressed rationally in $\La_j$, $\la_j$.
The full solution $\forall n$ was given by Babelon \cite{Bab2003,Bab2004} who obtained
it by analyzing the hyperelliptic spectral curve for the quasiperiodic chain.
The solution given below leads to the same final formulae as in \cite{Bab2003,Bab2004} but is simpler
since we work only with the rational spectral curve from the very beginning.

{\bf Solution.} Introduce two covectors:
\begin{subequations}\label{defphi}
\begin{alignat}{2}
    \Phi(u)&\equiv\bra{f}\Ucal=\bigl(\phi_1(u),\ldots,\phi_n(u)\bigr), &\qquad
    \phi_j(u)&=\Ucal_{1j},\\
    \Phit(u)&\equiv\bra{\ft}\Ucal=\bigl(\phit_1(u),\ldots,\phit_n(u)\bigr), &\qquad
    \phit_j(u)&=\re^{-x_n}\Ucal_{nj},
\end{alignat}
\end{subequations}
that provide an analog of Baker-Akhiezer function for two leaves of a rational Riemann surface
\cite{KV2000}.
Using $2\times2$ matrices we obtain:
\beq\label{phiAC}
    \phi_j(u)=A_{n,j+1}(u), \qquad \phit_j(u)=C_{j1}(u).
\eeq

Properties of $\Phi\Phit$:

\begin{enumerate}
\item $\Phi\Phit$ are polynomials in $u$ of degree $\leq n-1$.

\item Asymptotics as $u\rightarrow\infty$:
\begin{subequations}\label{asympt_phiphi}
\beq\label{asympt_phi}
    \phi_j(u)=u^{n-j}+u^{n-j-1}(X_n+\ldots+X_{j+1})+O(u^{n-j-2}),
\eeq
in particular, $\phi_n=1$. Also,
\beq\label{asympt_phit}
    \phit_j(u)=\re^{-x_j}u^{j-1}+O(u^{j-2}),
\eeq
in particular, $\phit_1=\re^{-x_1}$.
\end{subequations}
\item Conjugation conditions at points $u=\la_j$:
\beq\label{conjugphi}
    \Phit(\la_k)=\Phi(\la_k)\La_k, \qquad k=1,\ldots,n.
\eeq

The equations \Ref{conjugphi} are identical to Babelon's \cite{Bab2004} equations, 
though our interpretation in terms of the rational spectral curve is quite different.
\end{enumerate}

The properties (i) and (ii) are easily verified.
To derive the conjugation conditions (iii)
it is sufficient to show that the polynomials
\beq
   F_j(u)\equiv\phit_j(u)-\phi_j(u)C(u)
\eeq
are divisible by $A(u)$ $\forall j\in \{1,\ldots,n\}$. Using \Ref{phiAC}
and the shorthand notation
\beq
    L_{n,j+1}\equiv L''=\begin{pmatrix}A''&B''\\C''&D''\end{pmatrix},\quad
    L_{j1}\equiv L'=\begin{pmatrix}A'&B'\\C'&D'\end{pmatrix},\quad
    L_{n1}\equiv L=L''L',
\eeq
we obtain
\beq\label{def-F-cl}
    F_j(u)=C'-A''C.
\eeq

Multiplying the first term $C'$ by $1=\det L''=A''D''-B''C''$,
and substituting $C=C''A'+D''C'$ into the second term
one obtains, after, expanding the brackets,
\begin{align}
   F_j&=(A''D''-B''C'')C'-A''(C''A'+D''C')\notag\\
      &=-C''(B''C'+A''A')\notag\\
      &=-C''A.
\end{align}
\endproof

The same can be derived in terms of the matrix $\Ucal$.
By virtue of \Ref{4idents} and \Ref{defphi} we have
\beq
    F_j(u)=1\cdot\phit_j(u)-\phi_j(u)\cdot C(u)
    =\Ucal_{1n}\cdot\re^{-x_n}\Ucal_{nj}-\Ucal_{1j}\cdot\re^{-x_n}\Ucal_{nn},
\eeq
or
\beq
    -\re^{x_n}F_j=\Ucal_{1j}\Ucal_{nn}-\Ucal_{1n}\Ucal_{nj}
    =\begin{vmatrix}
       \Ucal_{1j} & \Ucal_{1n} \\
       \Ucal_{nj} & \Ucal_{nn}
    \end{vmatrix}.
\eeq

It remains to quote a theorem on the divisibility 
of the minors of the adjunct matrix by the determinant of the original
matrix \cite{Gan1966}
(Chap.\ 1, Sect.\ 4, formula (33)).

The properties (i--iii) are sufficient to reconstruct $\Phi\Phit$ from $\la\La$.
Indeed, introducing explicitly the coefficients of the polynomials
$\phi_,(u)$, $\phit_j(u)$
\begin{subequations}
\begin{align}
     \phi_j(u)&=r^{(j)}_0+r^{(j)}_1u+\ldots+r^{(j)}_{n-j-1}u^{n-j-1}+u^{n-j},\\
     \phit_j(u)&=q^{(j)}_0+q^{(j)}_1u+\ldots+q^{(j)}_{j-1}u^{j-1},
\end{align}
\end{subequations}
we obtain for them a system of equations that follows from the conjugation conditions
\Ref{conjugphi}
\beq
    q^{(j)}_0+\ldots+q^{(j)}_{j-1}\la_k^{j-1}
    -r^{(j)}_0\La_k-\ldots-r^{(j)}_{n-j-1}\la_k^{n-j-1}\La_k=\la_k^{n-j}\La_k,
\eeq
for $j=1,\ldots,n$.

The solution for $q^{(j)}_m$ and $r^{(j)}_m$ can be found by Cramer's rule.
Then the rational formulae for $\re^{-x_j}$ and $X_n+\ldots+X_{j+1}$ 
in terms of $(\La\la)$ can be extracted from the asymptotics \Ref{asympt_phiphi}.
See \cite{Bab2003,Bab2004} for the details.
The commuting quantities  $\re^{-x_j}$ provide thus the Hamiltonians for the 
dual (bispectral) problem.

\section{Quantisation}

Our treatment of the quantum case completely parallels the classical one.
It is more convenient to work with the $2\times2$ Lax matrix rather then with the
$n\times n$ one.

The Poisson brackets \Ref{eq:Pb} are replaced with the commutation relations
\beq
   [X_j,\re^{\pm x_k}]=\mp\ri\eta\re^{\pm x_k},
\eeq
the deformation parameter $\eta$ playing the role of the Planck constant.

The local Lax matrices $\ell_j(u)$ as well as the monodromy matrices $L_{jk}(u)$
and $L(u)$ are defined by the same formulae \Ref{eq:def-ell}, \Ref{defLjk},
and \Ref{eq:toda-Lll} as in the classical case.

The classical $r$-matrix relations \Ref{eq:Pb-ellell} and \Ref{eq:Pb-LL}
are replaced with the quantum ones
\cite{EKS16,EKS53}
\beq\label{Rellell}
    R(u_{12})\overset{1}{\ell}(u_1)\overset{2}{\ell}(u_2)
    =\overset{2}{\ell}(u_2)\overset{1}{\ell}(u_1)R(u_{12}),\quad
    R(u)=u+\ri\eta\Pcal_{12},
\eeq
\beq\label{RLL}
    R(u_{12})\overset{1}{L}_{jk}(u_1)\overset{2}{L}_{jk}(u_2)
    =\overset{2}{L}_{jk}(u_2)\overset{1}{L}_{jk}(u_1)R(u_{12}).
\eeq

Note also the quantum determinant relation
\beq\label{q-det}
   \text{q-det} T(u)\equiv A(u-\ri\eta)\,D(u)-C(u-\ri\eta)\,B(u)=1
\eeq

Since we work only with the open chain, all we need is the AC-subalgebra of \Ref{RLL}
\beq\label{AACC}
     [A(u_1),A(u_2)]=[C(u_1),C(u_2)]=0,
\eeq
\beq\label{AC}
    (u-v)A(u)C(v)+\ri\eta C(u)A(v) = (u-v+\ri\eta)C(v)A(u)
\eeq

Note that the above relations also hold for the partial monodromy matrix entries
$A_{jk}$ and $C_{jk}$.

Following \cite{EKS16} we define $\la_j$ as the zeroes of the self-commuting 
operator-valued polynomial $A(u)$:
\beq\label{def-lambdaq}
    A(\la_j)=0,\qquad j=1,\ldots,n.
\eeq

Define for the operator-valued polynomial $F(u)$ the
``substitution from the right'' as
\beq\label{def-subst-right}
   F(u)=\sum_{m=0}^p F_m u^m
   \quad\Longrightarrow\quad 
   \bigl[F(u)\bigr]_{u=\la_k}\equiv \sum_{m=0}^p F_m \la_k^m.
\eeq

Note that in \cite{EKS16} the ``substitution from the left'' is used instead
but it makes only a little change for the calculations.

Using \Ref{def-subst-right} define $\La_j$ as
\beq
  \La_k=\bigl[C(u)\bigr]_{u=\la_k},
\eeq

The quantum interpolation formulae are identical to the classical ones \Ref{AC-interpol}, one only
needs to preserve the exact ordering of the operators as shown in \Ref{C-interpol}

Using the same argument as in  \cite{EKS16} one derives from \Ref{AACC}
and \Ref{AC} the commutation relations
\begin{subequations}
\beq
    [\la_j,\la_k]=[\La_j,\La_k]=0,
\eeq
\beq\label{comm-La-la}
   \La_j\la_k=(\la_k+\ri\eta\d_{jk})\La_j
\eeq
\end{subequations}
replacing the classical Poisson brackets \Ref{pb-lala}.
For example, substituting $v=\la_j$ into \Ref{AC} from the right
one obtains
\beq
    A(u)\La_j(u-\la_j)=\La_j A(u)(u-\la_j+\ri\eta)
\eeq
whence \Ref{comm-La-la} follows after substituting \Ref{A-interpol}, like in  \cite{EKS16}.

To solve the inverse problem in the quantum case we define the covectors $\Phi\tilde\Phi$
by the same formulae \Ref{phiAC} as in the classical case. Then we verify the same properties (i)--(iii)
as in section 2. The properties (i)--(ii) being as trivial as in the classical case, we concentrate on proving
(iii). Note that the order of operators in the conjugation conditions   is now important,
so \Ref{conjugphi} is now replaced with
\beq
      \bigl[\tilde\phi_j(u)\bigr]_{u=\la_k}
      =\bigl[\phi_j(u)\bigr]_{u=\la_k}\La_k,\qquad\forall j,k
\eeq
or, using  $\tilde\phi_j(u)=C'(u)\equiv C_{j,1}(u)$ and
    $\phi_j(u)=A''(u)\equiv A_{n,j+1}(u)$,
\beq
  \bigl[C'(u)\bigr]_{u=\la_k}
      =\bigl[A''(u)\bigr]_{u=\la_k}\La_k,\qquad\forall j,k.
\eeq

Amazingly, the derivation for the classical case has to be only slightly modified to be adapted to the quantum case.

Replace the classical formula \Ref{def-F-cl} with
\beq\label{def-F-q}
    F_j(u)\equiv 1\cdot C'(u)-A''(u-\ri\eta)\cdot C(u).
\eeq

The proof of (iii) is then given by the following chain of equalities
\begin{align*}
  F_j(u)&=\bigl({A''(u-\ri\eta)\,D''(u)}-C''(u-\ri\eta)\,B''(u)\bigr)\cdot C'(u)\\
     &\phantom{=}-A''(u-\ri\eta)\cdot\bigl(C''(u)A'(u)+{D''(u)C'(u)}\bigr)\\
     &=-C''(u-\ri\eta)\,B''(u)\bigr)\,C'(u)-{A''(u-\ri\eta)C''(u)}\cdot A'(u)\\
     &=-C''(u-\ri\eta)\,A(u).
\end{align*}
using only the quantum determinant formula \Ref{q-det}
and the identity 
\beq
   A(u-\ri\eta)\,C(u)=C(u-\ri\eta)\,A(u)
\eeq
that follows from \Ref{AC} for $u-v=-\ri\eta$.

As soon as the properties (i)--(iii) are established, the rest is reduced to solving the systems
of linear equations for the coefficients of $\Phi\tilde\Phi$. For the details see \cite{Bab2003,Bab2004}.
The commuting quantities  $\re^{-x_j}$ are expressed as finite-difference operators in $\la_k$ with
rational coefficients, providing thus the Hamiltonians for the 
dual (bispectral) problem.

\section{Discussion}

Our improvement of the original derivation by \cite{Bab2003,Bab2004} is twofold. In the classical case,
we stress the use of the rational spectral curve, without recurse to the complicated hyperelliptic algebraic geometry.
In the quantum case, we rely solely on the $R$-matrix algebra \Ref{RLL}, without recurse to the 
integral representations for the eigenfunctions. As a consequence, our approach must also work  for other
integrable models of $R$-matrix type, like relativistic Toda chain, or XXX and XXZ spin chains.

Our approach to solving the inverse problem can also be used in the analysis of correlation functions
for the Toda lattice \cite{Koz2013}.

\section*{Acknowledgements}

The work is supported by EPSRC grant EP/H000054/1.
I am grateful to Karol Koz{\l}owski for discussions and for letting me know of his article \cite{Koz2013}.


\end{document}